# Tetrahedral Magnetic Order and the Metal-Insulator Transition in the Pyrochlore Lattice of $Cd_2Os_2O_7$

J. Yamaura,[1] K. Ohgushi,[1] H. Ohsumi,[2] T. Hasegawa,[3] I. Yamauchi,[1] K. Sugimoto,[4] S. Takeshita,[2] A. Tokuda,[2,5] M. Takata,[2,4] M. Udagawa,[3] M. Takigawa,[1] H. Harima,[6] T. Arima,[2,7] and Z. Hiroi[1]

[1]*Institute for Solid State Physics, University of Tokyo, Kashiwa, Chiba 277-8581, Japan*
[2]*RIKEN, SPring-8 Center, Sayo, Hyogo 679-5148, Japan*
[3]*Graduate School of Integrated Arts & Sciences, Hiroshima University, Higashi-Hiroshima, Hiroshima 739-8521, Japan*
[4]*Japan Synchrotron Radiation Research Institute, SPring-8, Sayo, Hyogo 679-5198, Japan*
[5]*Department of Physics, Kwansei Gakuin University, Sanda, Hyogo 669-1337, Japan*
[6]*Department of Physics, Graduate School of Science, Kobe University, Kobe, Hyogo 657-8501, Japan*
[7]*Department of Advanced Materials, University of Tokyo, Kashiwa, Chiba 277-8561, Japan*
(Received 26 November 2011)

$Cd_2Os_2O_7$ shows a peculiar metal-insulator transition at 227 K with magnetic ordering in a frustrated pyrochlore lattice, but its magnetic structure in the ordered state and the transition origin are yet uncovered. We observed a commensurate magnetic peak by resonant x-ray scattering in a high-quality single crystal. X-ray diffraction and Raman scattering experiments confirmed that the transition is not accompanied with any spatial symmetry breaking. We propose a noncollinear all-in/all-out spin arrangement on the tetrahedral network made of Os atoms. Based on this we suggest that the transition is not caused by Slater mechanism as believed earlier but by an alternative mechanism related to the formation of the specific tetrahedral magnetic order on the pyrochlore lattice in the presence of strong spin-orbit interactions.

The metal-insulator (MI) transition is one of the most dramatic phenomena observed for electrons in solids. Generally, a strong electron correlation or Fermi surface instability causes a solid to transition from showing conductive behavior at high temperature to showing localized insulator behavior at low temperature. The transition caused by the strong electron-electron interaction is known as the Mott transition, which is of the first order and in many cases is accompanied by a simple collinear-type antiferromagnetic ordered state [1]. A prime example of the MI transition caused by Fermi surface instability is the charge- or spin-density wave (CDW or SDW) transition; in this case, the magnetic structures exist in a nonmagnetic singlet state or an incommensurate state. In addition, in weak coupling systems, another type of MI transition—the Slater transition—is theoretically expected to occur via folding of the Brillouin zone owing to the antiferromagnetic order [2]. Given that frustration typically inhibits the antiferromagnetic order, previously documented MI transitions with magnetic ordering are expected to be less probable on frustrated lattices. Consequently, an MI transition due to new mechanism leading to a peculiar ground state that eliminates the frustration is expected.

One of the most extensively researched materials is the pyrochlore oxide [3]. Two classes of pyrochlore oxides with MI transitions have been identified: one class exhibits transitions with a large structural distortion and the other class superficially shows no distortion. The former class includes $Tl_2Ru_2O_7$[4,5], and $Hg_2Ru_2O_7$ [6-8], in which the singlet state or magnetic long-range order appears with the relaxation of frustrations owing to the large lattice distortions. The latter class includes $Cd_2Os_2O_7$ [9,10] and $R_2Ir_2O_7$ (R = Nd, Sm, Eu) [11], in which certain magnetic orders seem to occur simultaneously at the MI transition; they have not yet been determined experimentally. It is a mystery why the magnetic orders could emerge with frustration surviving, if the pyrochlore lattice remains undistorted.

The MI transition of $Cd_2Os_2O_7$ at 225 K was discovered by Sleight in 1974 [9]. Subsequently, Mandrus *et al.* [10] and Padilla *et al.* [12] pointed out the Slater transition as the origin of this MI transition. Recently, Koda *et al.* found a static internal magnetic field and suggested the appearance of an incommensurate SDW transition [13]. However, Singh *et al.* disputed the appearance of the Slater transition owing to the absence of a nesting feature in the band structures [14]. Reading and Weller carried out neutron diffraction experiments using $^{114}$Cd-enriched polycrystalline samples; natural Cd is a neutron absorber. They did not find any magnetic peaks to identify the magnetic structure in the insulating phase. They found a discontinuity in the lattice parameters at the transition [15], which seems contradictory to other results of a second-order-type transition [10,12]. Further systematic study using a single crystal and by various methods are required to determine the magnetic structure and examine the possibility of structure transition. Recently, Chern and Batista predicted in their double-exchange models that a $q = 0$ magnetic order is stable and also that a gap opens by a Hund's coupling, assuming a tetrahedral magnetic order [16]. However, their band structures are too simple and substantially different from those calculated taking into account a strong spin-orbit (SO) interaction [14,17,18].

In this study, we successfully prepared high-quality single crystals of $Cd_2Os_2O_7$. We employed the resonant x-ray scattering (RXS) method, instead of the neutron scattering method which suffers from the absorption problem, to reveal the magnetic structure. The effectiveness of this technique in $5d$ transition metal compounds was recently demonstrated for $Sr_2IrO_4$ [19]. Moreover, Raman scattering experiments were performed. Consequently, we show that the MI transition is caused by a unique mechanism related to a specific spin arrangement on the pyrochlore lattice and not the Slater transition as proposed previously.

Single crystals of $Cd_2Os_2O_7$ were grown from CdO and Os in a sealed quartz tube, under supply of an appropriate amount of AgO as the oxygen source in a temperature gradient furnace maintained at 900–750 °C for a week (Fig. 1, inset). RXS was carried out at the absorption edge of Os $L3$ ($E$ = 10.874 keV) by using a six-circle diffractometer at the BL19LXU of SPring-8 [20]. The experimental setup is schematically depicted in the inset of Fig. 2. RXS was also carried out at the absorption edge of Os $L2$ ($E$ = 12.387 keV) to detect magnetic diffraction peaks in a wide $2\theta$-range, by using an imaging plate at the BL02B1 of SPring-8 [21]. Raman scattering spectra were measured using a liquid-$N_2$-cooled CCD detector along with a triple-monochromator. The 514.5-nm line of an Ar laser was employed as excitation light.

First, we carefully investigated the possibility of lattice distortion. In the high-resolution x-ray experiment using synchrotron radiation at 100 K, observation of many fundamental reflections at $2\theta$ = 90–110° revealed neither peak splitting nor broadening of the reflection via distortion of the lattice from the cubic system. This indicates that the lattice is cubic within an error of $\Delta a/a$ = 0.0004. Moreover, in the $00l$ ($1 \leq l \leq 10$) and $0kl$ ($2 \leq k, l \leq 6$) scans and observation of the oscillation photographs on the imaging plate, neither additional reflection breaking the face-centered lattice nor long-period structure was observed. In contrast, the $00l$ ($l = 4n + 2$) and $0kl$ ($k+l = 4n + 2$) reflections were observed below $T_{MI}$, indicating violation of the $Fd$-$3m$ extinction rule. However, as will be discussed later, the observed additional reflections are attributable not to lattice distortion but to the appearance of magnetic scattering. The results imply that the space group in the low-temperature phase is the same as that in the high-temperature phase, $Fd$-$3m$.

The preservation of the spatial symmetry was more reliably confirmed by a measurement of Raman scattering, which is the most powerful technique for detecting small lattice distortions. Raman spectra at 300 and 4 K are shown in Fig. 1. Six Raman active modes of $A_{1g} + E_g + 4T_{2g}$ in the parallel polarization geometry, observed at both temperatures, were completely consistent with $Fd$-$3m$ [22]. Consequently, we conclude that spatial symmetry breaking does not occur with the MI transition in this compound.

RXS is effective for observing very weak x-ray magnetic reflections even from a tiny crystal. However, it is necessary to distinguish two contributions from crystallographic and magnetic scattering. The anisotropy in the atomic scattering factors, arising from the local atomic environment, shows a resonant enhancement at the absorption edges and hence may break the extinction rule. This is referred to as anisotropic-tensor-susceptibility (ATS) scattering [23]. On the basis of the crystal symmetry as treated by Dmitrienko [24], the tensorial structure factor of $00l$ ($l = 4n + 2$) reflections due to the ATS scattering, which are forbidden by the space group $Fd$-$3m$, is evaluated as three by three matrix $\boldsymbol{F}_{00l}$ = {0, $F_{ATS}$, 0, $F_{ATS}$, 0, 0, 0, 0, 0}. Then, the ATS intensities of a $00l$ reflection at the azimuthal angle $\psi$ in the π-σ′ and π-π′ polarization channels are calculated to be $I_{00l}$ = $F_{ATS}^2\sin^2\theta\cos^2 2\psi$ and $F_{ATS}^2\sin^4\theta\sin^2 2\psi$, respectively. Hence, the contribution of the ATS scattering to the $00l$ ($l$ = 4n+2) reflection is negligible at $\psi$ = ±45° in the π-σ′ channel and at $\psi$ = 0° and 90° in the π-π′ channel.

Figure 2 shows the spectra near the Os $L3$ edge of x-ray absorption and the intensity of the 006 reflection at 10 K. The 006 reflection was obtained in the π-σ′ channel at $\psi$= -47°; thus, the contribution of the ATS scattering must be vanishingly low. The maximum of the 006 reflection is in good agreement with the Os $L3$ edge, as depicted by the dotted line, implying an enhancement by the resonance effect. The small peak marked by M is induced by multiple scattering. The left inset of Fig. 2 shows the polarization dependence of the 006 reflection at 10 K without the ATS contribution. The intensity in the π-σ′ channel at $\psi$ = -47 ° is 20 times as large as that in the π-π′ channel at $\psi$ = 0°; the weak intensity in the π-π′ channel is ascribed to a leakage component from the π-σ′ channel, which can appear because the $2\theta$ value of the polarizer is not precisely 90°. Thus, the 006 reflection appears in only the π-σ′ channel, indicating magnetic scattering in origin. We therefore conclude that a commensurate magnetic structure appears at the propagation vector $q$ = 0 at low temperature.

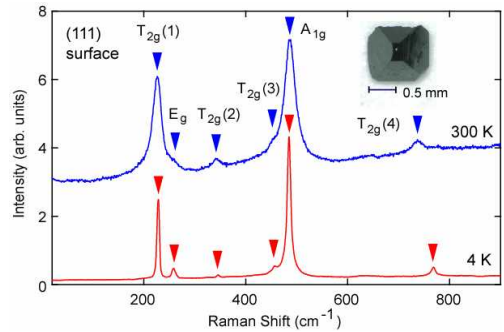

FIG. 1. (Color online) Raman spectra at 300 and 4 K on the (111) surface of a $Cd_2Os_2O_7$ single crystal in parallel polarization. The inset shows a single crystal of $Cd_2Os_2O_7$.

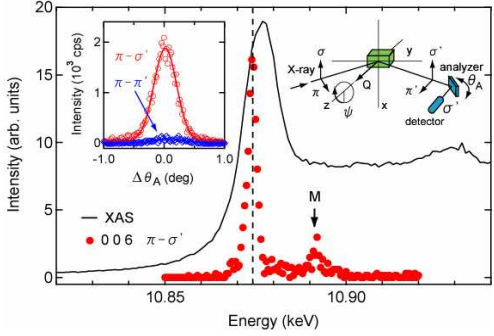

FIG. 2. (Color online)Spectra of the intensity of the 006 reflection in the π-σ′ channel ($\psi$= -47°) at 10 K and X-ray absorption near the Os $L3$ edge at room temperature. M represents the multiple scattering. The right inset shows the experimental setting in the π-σ′ channel, where the azimuthal angle $\psi$ (angle in the in-plane rotation of the crystal) is defined to be zero when the $x$-axis is parallel to the $a$-axis of the crystal. The left inset shows the polarization dependence of the 006 reflection at 10.874 keV.

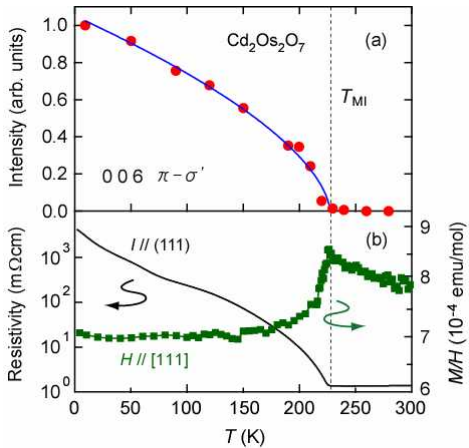

FIG. 3. (Color online) (a) Temperature dependence of the integrated intensity of the 006 magnetic reflection measured in the π-σ′ channel at $\psi$ = -47°. Upon cooling, it increases gradually below $T_{MI}$ = 227 K following a power law of temperature, indicating a second-order transition. (b) Resistivity in a zero magnetic field and magnetic susceptibility in $H$ = 20 kOe measured after zero-field cooling, both of which show clear anomalies at $T_{MI}$.

The 006 reflection gains its intensity gradually below $T_{MI}$ = 227 K, as shown in Fig. 3, indicating a continuous second-order transition. It is obvious that the 006 reflection is magnetic and induced by the MI transition, because anomalies in both magnetic susceptibility and resistivity are observed at nearly the same temperature.

To understand the mechanism of MI transition, it is important to know the spin arrangement in the magnetic ordered phase. The $q$ = 0 magnetic structure should be described in terms of an Os tetrahedron because the magnetic primitive cell includes one Os tetrahedron. Here, the possible spin arrangements of the Os site can be classified by the representation analysis technique; the magnetic representation is expressed as $\Gamma_{Os} = 1\Gamma_3^{(1)}+1\Gamma_5^{(2)}+1\Gamma_7^{(3)}+2\Gamma_9^{(3)}$, where the superscript denotes the number of the basis functions in the irreducible representations [25]. The spin arrangements of the 12-basis functions are given in Ref. 26, and those of five selected basis functions $\Psi_1$, $\Psi_2$, $\Psi_3$, $\Psi_4$, and $\Psi_8$ are illustrated in Fig. 4. Here, we calculate the magnetic structure factor with the aim of rationalizing the observed polarization dependence. The simplified magnetic structure factor is written as $f_{mag} = F\sum_j(\varepsilon_i \times \varepsilon_s) \cdot m_j e^{2\pi q r_j}$, which is a predominant term for the magnetic scattering, where $F$ is the difference of the transition strengths with the helicity, $m_j$ is the spin vector at each Os site, $q$ is the propagation vector, $r_j$ is the position of the Os atom, and $\varepsilon_i$ and $\varepsilon_s$ are the polarization unit vectors for the incident and scattered x-rays, respectively [27]. Table I lists the calculated intensities of the 006 reflection for the basis functions in the π-σ′ channel at $\psi$ = -45° (experimentally observed) and in the π-π′ channel at $\psi$ = 0° (unobserved). A qualitative analysis of the experimental and calculation results leads to the following conclusions. First, $\Psi_6$ and $\Psi_{10}$ do not contribute to the formation of the magnetic structure. Second, $\Psi_1$, $\Psi_2$, $\Psi_3$, $\Psi_4$, and $\Psi_8$ can independently describe the magnetic structure. Third, a linear combination of functions within the same irreducible representation, such as $\Psi_2+c\Psi_3$, $\Psi_4+c\Psi_5$, and $\Psi_8+c\Psi_{12}$, can also account for the magnetic structure because the transition is of the second order. All of these combinations are noncollinear antiferromagnetic spin arrangements. The ferromagnetic spin arrangements $\Psi_7$, $\Psi_9$, and $\Psi_{11}$ are excluded because of the absence of large magnetization below $T_{MI}$ as shown in Fig. 3(b).

The magnetic structure cannot be uniquely determined by these results alone. It can, however, be deduced from the lattice and magnetic symmetries. Generally, in a transition metal oxide, the lattice and the magnetic structure are strongly coupled through the exchange striction and the SO interaction. Therefore, a low-symmetry spin alignment almost always induces a crystallographic distortion. For example, the spinel oxide ZnCr$_2$O$_4$ shows a spin-driven Jahn-Teller distortion [28]. Further, the pyrochlore oxides Tl$_2$Ru$_2$O$_7$ and Hg$_2$Ru$_2$O$_7$ also exhibit large lattice distortions with the MI transition [4–8]. Thus, the magnetic symmetry is expected to be associated with the lattice symmetry in Cd$_2$Os$_2$O$_7$ as well.

In the antiferromagnetic basis functions, only $\Psi_1$ indicates cubic magnetic symmetry with just one magnetic interaction, whereas the other basis functions indicate tetragonal magnetic symmetry with two magnetic

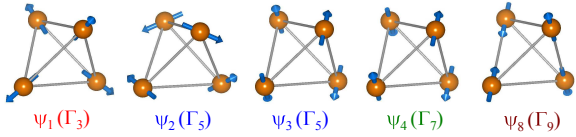

$\psi_1\,(\Gamma_3)$ $\quad$ $\psi_2\,(\Gamma_5)$ $\quad$ $\psi_3\,(\Gamma_5)$ $\quad$ $\psi_4\,(\Gamma_7)$ $\quad$ $\psi_8\,(\Gamma_9)$

FIG. 4. (Color online) Spin arrangements on a regular tetrahedron for selected basis functions $\Psi_i$ with $i = 1, 2, 3, 4$, and 8 and the corresponding irreducible representations $\Gamma_s$. All of them are noncollinear antiferromagnetic spin arrangements. Only the all-in/all-out $\Psi_1$ indicates cubic magnetic symmetry; the other antiferromagnetic basis functions indicate tetragonal magnetic symmetry.

TABLE I. Intensity of the 006 magnetic reflection, $f_{mag}^2$, in the $\pi$-$\sigma'$ channel at $\psi = -45°$ (experimentally observed) and in the $\pi$-$\pi'$ channel at $\psi = 0°$ (unobserved). A and B denote $2\sin\theta$ and $2\cos\theta$, respectively. All intensities are normalized by $F^2$. Ferromagnetic basis functions $\Psi_7$, $\Psi_9$, and $\Psi_{11}$ are omitted.

| | $\Gamma_3$ | $\Gamma_5$ | | $\Gamma_7$ | | | $\Gamma_9$ | | |
|---|---|---|---|---|---|---|---|---|---|
| | $\Psi_1$ | $\Psi_2$ | $\Psi_3$ | $\Psi_4$ | $\Psi_5$ | $\Psi_6$ | $\Psi_8$ | $\Psi_{10}$ | $\Psi_{12}$ |
| $\pi$-$\sigma'$ | $4/3A^2$ | $2/3A^2$ | $2A^2$ | $B^2$ | 0 | $B^2$ | $B^2$ | $B^2$ | 0 |
| $\pi$-$\pi'$ | 0 | 0 | 0 | 0 | 0 | $2A^2B^2$ | 0 | $2A^2B^2$ | 0 |

interactions between the Os spins. Provided that the latter functions contribute to the magnetic structure, a slight but detectable lattice distortion should be observed in the ordered phase; however, we did not observe any lattice distortion. Consequently, the $\Psi_1$ spin arrangement is a unique solution to the magnetic ordered structure. $\Psi_1$ denotes an all-in/all-out antiferromagnetic spin arrangement with a zero net moment on the Os tetrahedron, in which all the spins point to (all-in) or away (all-out) from the centers of the tetrahedron. This type of tetrahedral magnetic order has not been found except for a magnetic insulator FeF$_3$ [29].

The all-in/all-out order with $q = 0$ obtained in the present study seems to contradict the incommensurate SDW order proposed in the previous μSR experiments using a powder sample; the muon spin relaxation rate starts to increase at $T_{MI}$ and shows a peak at ~150 K, below which static magnetic moments are observed [13]. This apparent inconsistency may be partly due to the difference in sample quality between powder and single crystals. Nevertheless, there might be another possibility. The SDW order was suggested on the basis of the observation of two internal magnetic fields of different magnitude, one of which was nearly zero, and by assuming that there was a single site for implanted muons in the crystal. If there are multiple muon sites, however, alternative interpretation would become conceivable. Presumably, there are some muon sites where the internal field from Os moments in the all-in/all-out arrangement tends to be cancelled partially or completely due to the geometrical reason. If this is the case, the muon depolarization could not detect large longitudinal but only small transverse fluctuating components. Hence, this seemingly inconsistent conclusion may come from the difference in experimental probes. In order to clarify this issue as well as other mysteries left in the interpretation of the previous μSR observations, we plan to carry out further μSR experiments using high-quality single crystals.

Finally, we briefly discuss the origin of the MI transition in Cd$_2$Os$_2$O$_7$. Considering the on-site Coulomb energy lying in the same order of the band width, Mandrus *et al.* suggested that the MI transition of this compound is of the Slater type (*i.e.*, a continuous phase transition in the weak coupling region) rather than of the Mott type [10]. However, the present spin arrangement with $q = 0$ indicates that this transition is not due to Slater-type or SDW transitions. Consequently, the band-gap formation of this compound is thought to arise from an alternative mechanism that is presumably related to the all-in/all-out spin arrangement: a tetrahedral magnetic structure on the pyrochlore lattice without any spatial symmetry breaking.

Let us consider a possible mechanism to realize an MI transition without spatial symmetry breaking. According to band structure calculations [14,17], the nonmagnetic phase of Cd$_2$Os$_2$O$_7$ is a semimetal with small electron and hole pockets at the Fermi energy, which is consistent with the metallic behavior of the high-temperature phase. Thus, an MI transition without spatial symmetry breaking can occur when the hole and electron bands move upward and downward, respectively, and finally lose their overlap; it may be considered as a Lifshitz transition. However, a large band shift by ~40 meV is required for Cd$_2$Os$_2$O$_7$ [17]. Moreover, because the band structure of nonmagnetic Cd$_2$Os$_2$O$_7$ near the Fermi energy consists of forty 5$d$ orbitals from four Os atoms in the primitive cell, such simple band shifts are not likely to take place without a specific reason. We think that the emergence of a tetrahedral magnetic order, that is found in the low-temperature phase, triggers such cooperative band shifts so as to form a small gap through a strong SO coupling. Future band structure calculations taking into account the tetrahedral magnetic order and SO interactions would confirm and clarify this novel mechanism. Strong SO interactions may also play a crucial role in selecting the all-in/all-out spin arrangement on the pyrochlore lattice of Cd$_2$Os$_2$O$_7$ [30].

In summary, we found a commensurate magnetic peak with the propagation vector $q = 0$ below the MI transition of Cd$_2$Os$_2$O$_7$ by resonant x-ray scattering measurements. Discussion in terms of an irreducible representation reveals that the ordered magnetic structure is most likely to be a noncollinear all-in/all-out spin arrangement with a zero net moment on the Os tetrahedron. We suggest that the symmetry of the pyrochlore lattice is notably apparent in the spin arrangement and insulating mechanism of this compound.

We thank H. Shinaoka for helpful discussion. The synchrotron radiation experiments were performed at beamlines BL-02B1 and BL-19XU at SPring-8 with the approval of RIKEN (Proposal Nos. 2009B1200 and 20090064). This work was supported by a Grant-in-Aid for Scientific Research on Priority Areas "Novel States of Matter Induced by Frustration" (19052003).